**Title:** Disordered But Rhythmic: the role of intrinsic protein disorder in eukaryotic circadian timing

**Authors:** Emery T. Usher[1] and Jacqueline F. Pelham[2*]

**Affiliation**

[1.] Department of Chemistry and Biochemistry, University of California, Santa Cruz, Santa Cruz, CA 95064

[2] Department of Biochemistry and Molecular Biophysics, Washington University School of Medicine, St. Louis, MO 63116

**Correspondance**

*J. Pelham

Email: pelham@wustl.edu

## Abstract

Intrinsically disordered protein regions (IDRs) are found across all domains of life and are characterized by a lack of stable 3D structure. Nevertheless, IDRs play critical roles in the most tightly regulated cellular processes, including in the core circadian clock. The molecular oscillator at the heart of circadian regulation leverages IDRs as dynamic interaction modules–for activation and repression, alike–to support robust timekeeping and expand clock output and regulation. Here, we cover the biophysical mechanisms conferred by IDRs and their modulators. We survey the intrinsically disordered regions in clock proteins that are widely prevalent from fungi to mammals and discuss the importance of IDRs to the core clock and beyond.

## Introduction

The planet's continuous 24-hr light-dark cycle has led to the evolution of timing mechanisms that coordinate the internal physiology of organisms with their environment. These approximately 24-hr biological oscillations, referred to as circadian rhythms (*circa* - "about" and *dies* - "day"), are found in all domains of life and regulate many aspects of physiology [1–5]. A molecular clock coordinates the circadian system, integrating signals such as light and temperature to synchronize internal physiology and behavior with the environment.

Coordination with the external environment provides a competitive advantage across the tree of life, from predation to the timing of photosynthesis to reproductive efforts [6–8]. Due to the breadth of different environments, biological niches, and physiological needs, circadian regulation is context-dependent and varies accordingly. Although there is vast variation amongst organisms, it is still possible to capitalize on the power of model systems, as many features of timing are conserved or converged upon and shared across the kingdoms of life [9–12].

The field employs several robust systems to dissect the molecular bases of circadian timing. *Neurospora crassa* and *Drosophila melanogaster* have been used to make many seminal discoveries of eukaryotic timing. *Neurospora* is a filamentous fungus that is genetically tractable and has provided conceptual framework for animal clocks, including oscillator architecture, light and clock input, and many of the other clock-controlled biological processes. [10,13–15]. *Drosophila* was the first organism to have a clock gene cloned, and has since made many contributions to the field, including clock mechanisms, environmental regulation, circadian neurobiology, sleep, and circadian ties to reproduction [16–20].

The dissection of the mammalian central pacemaker (the Suprachiasmatic Nucleus in the brain) and the molecular oscillator was facilitated by genetic and behavioral screens. Once the genetic revolution occurred and molecular cloning became tractable in mammals, *Mus musculus* enabled the identification of the first mammalian clock gene [23,35,36]. Vertebrate studies led to many advancements, including the association of the clock with human diseases, and have since expanded to uncover genome-wide regulation by the clock [37–40]. Despite the considerable differences between these taxa, the architecture of the circadian molecular oscillator is generally conserved (Box 1).

**Box 1: The molecular bases of fungal and animal circadian timing: a TTFL view**

In fungi and animals, the autonomous TTFL is a feedback loop driven by negative repression [2]. The primary feedback loop drivers are a set of interconnected activating and repressing proteins. The cycle begins with the transcriptional activation phase, in which heterodimeric transcription factors bind to E-boxes or Light-responsive elements (LREs, C-box) within the promoters of the negative-repressing elements [21–24] (Fig. 1). The activators in the fungal clock are WC-1 and WC-2, which, upon dimerization, form the White Collar Complex (WCC) [25]. The mammalian heterodimer includes Brain and Muscle ARNT-Like 1 (BMAL1, also known as ARNTL) and Circadian Locomotor Output Cycles Protein Kaput (CLOCK), and the *Drosophila* counterparts are Circadian Locomotor Output Cycles Kaput (dCLK) and CYCLE (CYC) (Fig. 1).

Activator dimerization and DNA binding leads to transcription of negative element proteins. Once translated, the repressing elements assemble with other proteins, migrate back to the nucleus, and directly inhibit their own transcription (Fig. 1). In mammals, BMAL1:CLOCK activate the repressor genes cryptochrome (*Cry*) and period (*Per)* (encoding paralogs CRY1, CRY2, PER1, PER2, and PER3). The functional analogs activated by the WCC in *Neurospora* are FREQUENCY (FRQ) and FREQUENCY-Interacting RNA Helicase (FRH). *Drosophila* repressors include PERIOD (dPER) and TIMELESS (TIM). A unifying feature of the repressing complex in both fungi and animals is casein kinase 1 (CK1) (or ortholog, DBT in *Drosophila*) as a constituent.

Progressive phosphorylation of the negative elements by CK1 and other kinases leads to their degradation, which in turn lifts inhibition and allows the cycle to restart [26–29]. Several critical delays within the oscillator are essential for generating sustained rhythms [27,30–34]. Another crucial feature for the stability of the oscillator mechanism is the precise control of stability and spatiotemporal regulation of the core clock proteins(Fig. 1).

Core clock proteins share a common set of conserved structural and functional elements essential for clock function. PAS (Per-Arnt-Sim) domains serve as dimerization domains for both clock activators and repressors in animals and fungi. The clock activators also all have DNA binding domains for their transcriptional activation roles (expanded on in Box 2). It is important to note that the proteins and mechanisms mentioned here are not an exhaustive list, as the complexity of these model systems is considerable.

The core eukaryotic molecular oscillator, referred to here as the circadian clock, is a transcription-translation feedback loop (TTFL). The underlying mechanism associated with the TTFL is autoregulatory negative repression (Box 1 and Fig. 1). The TTFL times the expression of numerous clock-controlled genes (*ccgs*) [4,13,41] (Fig. 1), which causes temporal oscillations in many biological processes, known collectively as 'clock output'. Furthermore, there is extensive circadian regulation that occurs at the post-transcriptional and post-translational levels [39,42–45].

There is conservation of the general molecular architecture of the TTFL, as well as conservation of structured protein domains (Boxes 1 and 2). The field has made remarkable progress in determining the mechanisms and functions associated with the structured regions of the clock proteins (Box 2). However, we still have a comparatively limited understanding of the mechanistic functions associated with the regions without structure, the intrinsically disordered regions (IDRs). Considering that IDRs represent over half of the sequence space of the key core eukaryotic clock proteins; our models of the TTFL are incomplete without roles and mechanisms for these flexible regions [46].

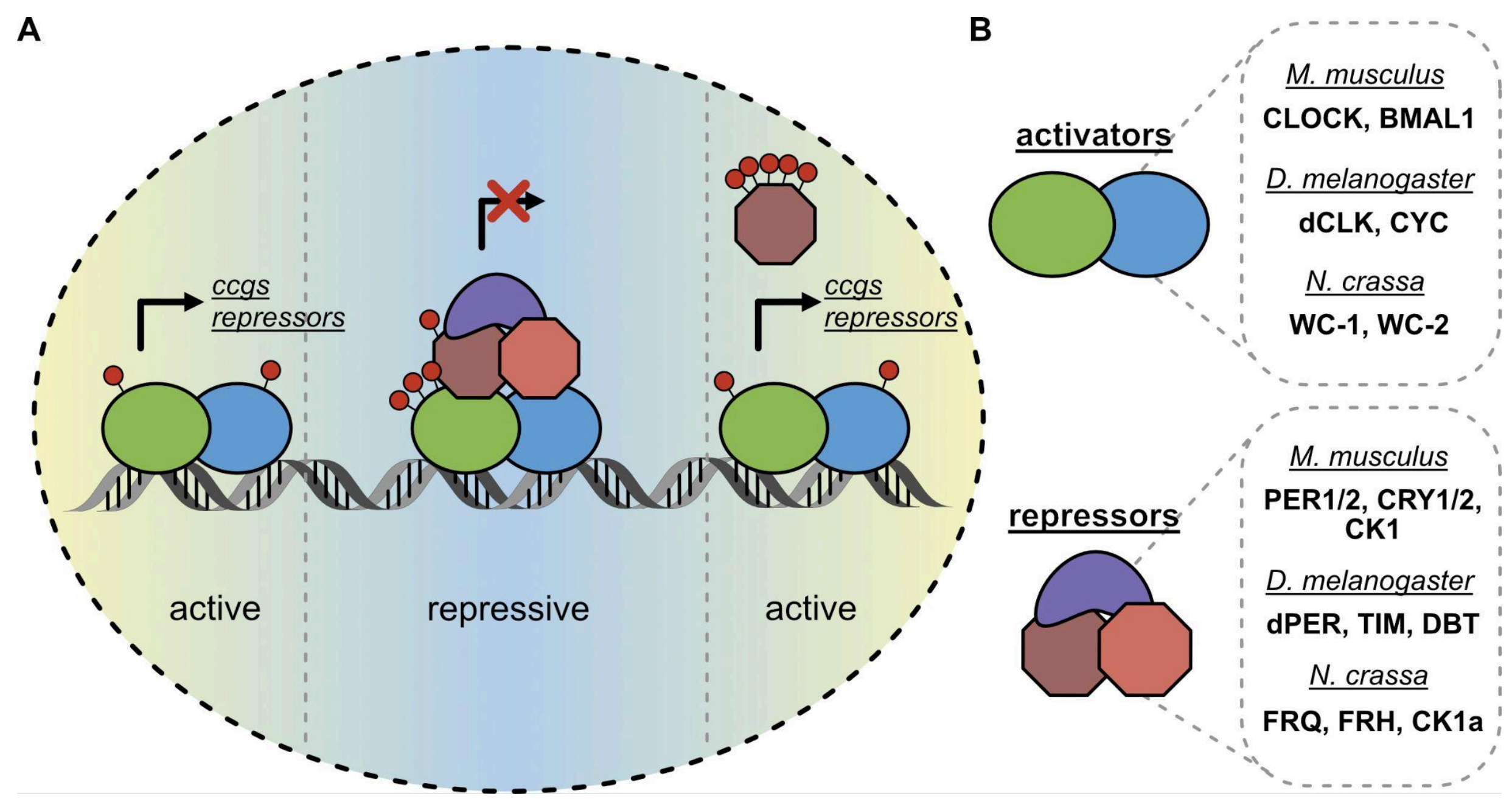

**Figure 1: TTFL architecture across different eukaryotes.** (A) Simplified schematic of the active and repressive phases of the circadian timekeeping system depicting five key players in the eukaryotic clock (detailed in Box 1).Red circles represent phosphorylation marks. (B) Functional analogs of the activator and repressor proteins in *Mus musculus, Drosophila melanogaster,* and *Neurospora crassa*.

IDRs lack a stable 3D structure and instead occupy a heterogeneous collection of conformational states (reviewed in [47]). Their high degree of plasticity allows them to adapt to external conditions, such as the presence of binding partners, properties of the solution, and environmental conditions [48–52] (Figs. 2-3). Furthermore, standard structural determination techniques are not suited to investigating IDRs (Box 2). This has led to a large gap in knowledge surrounding the biophysical and mechanistic contributions of IDRs in circadian clock proteins [46].

Here, we briefly cover the molecular TTFL and the fundamentals of IDRs and their general biological and biophysical mechanisms. We also discuss examples of disorder-based timing mechanisms and processes. Lastly, we elaborate on the functional implications of disorder in and beyond the core oscillator.

## Fundamentals of intrinsically disordered protein regions

### *Characteristics of IDRs*

IDRs are defined by a lack of persistent secondary and tertiary structure such that–in general–they tend to be more expanded and more dynamic than their structured counterparts. In contrast to folded proteins, IDRs do not have a single lowest-energy conformation. Instead, IDRs are highly flexible and can interconvert between myriad conformational states; this collection of positions is known as the 'conformational ensemble' (Fig. 2A). IDRs are found across all domains of life in various contexts: some are true intrinsically disordered proteins (IDPs), which lack any regions of stable structure (Fig. 2A) [53–55]. Many IDRs, however, exist as linkers–connecting two folded domains, tails–extending toward the N- or C-terminus, or loops–spanning folded elements within a single domain (Fig. 2B).

Like globular domains, IDR sequence composition gives rise to specific structural features. IDRs are often depleted in the hydrophobic residues (C, F, I, L, V, W, and Y) that tend to drive protein folding and are enriched in polar and charged residues (D, E, K, Q, R, S, and T) [56,57]. Solvation of these hydrophilic amino acids supports the expanded dimensions of many IDRs. Glycine and proline residues are also abundant in IDR sequences, owing to their flexibility and rigidity, respectively [47][58]. Another hallmark of some IDRs is sequence repetition or 'low-complexity' regions (LCRs). LCRs are stretches of a sequence enriched in a few certain amino acids and depleted in most others, such as the heptapeptide repeats in the RNA Pol II C-terminal IDR or the pathologic polyglutamine (polyQ) tracts in huntingtin [59,60].

Although IDRs are largely unstructured, their conformational ensembles are almost *never* truly 'random'. The features of a conformational ensemble are dependent on both the physiochemical properties of the amino acid sequence and the context of the local solution environment (Fig. 3) [49,61]. IDR properties can be described as 'global'–features of the entire IDR ensemble, and 'local'–features of a subsection of the IDR (Fig. 2C-D). Global features include the expandedness or compactness of an IDR ensemble, whereas local features are those, such as α-helical propensity.

An emerging hypothesis is that IDRs contain transient secondary structure or engage in long-range intramolecular interactions that bias the conformational ensemble toward particular functions (Fig. 2D and 3) [62–66]. Depending on their number and relative locations, charged residues (D, E, K, R, and H, depending on pH) can bias an ensemble via electrostatic attraction or repulsion (Fig. 3A). For example, charge patterning–the relative location of positive and negative residues in an IDR

**Glossary**

*Conformational Ensemble:* the collection of poses that are energetically accessible to an IDR

*Transient structure:* an element of secondary structure (i.e., α-helix) that samples both helical and non-helical (disordered) conformations

*Radius of gyration:* a measure of ensemble dimensions; the average distance of each atom in the chain to the center of mass of the IDR

*End-to-end distance:* a measure of ensemble dimensions; the through-space distance between the N- and C-termini of an IDR

*Temperature compensation:* a feature that allows the clock to maintain a consistent speed despite temperature changes

*Short Linear Motif (SLiM):* amino acid sequence in an IDR, 3-12 residues long with conserved sequence elements that mediates protein-protein interactions

*Entrainment:* synchronization of one cycle with another, circadian entrainment synchronizes the physiological rhythm with the external 24 hr cycle

sequence–has been shown to influence the conformational behavior and function of transcriptional regulators [67].

Other attractive forces, like cation-π or hydrophobic interactions, can also contribute to the ensemble behavior. Hence, changing solution conditions can modulate IDR ensemble properties in a sequence-dependent manner (Fig. 3) [68]. Changing pH or salt concentration may modulate electrostatic interactions toward ensemble expansion or compaction [61,69,70] (Fig. 3A). Likewise, increasing temperature can affect chain solvation or weaken intramolecular interactions to drive ensemble expansion or compaction [71]. Molecular crowding–either via artificial crowders (e.g., PEG, dextran), or cellular crowding–can expand or compact IDRs in complex ways that depend on the IDR sequence and the crowder. (Fig. 3A) [49,72,73].

IDRs are also overrepresented as targets of post-translational modifications (PTMs) [74]; different PTMs (e.g., phosphorylation, introducing a negative charge) can change the *apparent* sequence chemistry to modulate ensemble properties and direct protein function [75,76] (Fig. 3B). Furthermore, PTMs that modify the same residue type can potentially compete with one another. For instance, the competition between acetylation and ubiquitination in the C-terminal IDR of p53 tunes its half-life [77]. Separately, existing PTMs may influence the placement of additional PTMs, such as the requirement for a 'priming' phosphate for certain kinases to add additional phosphates [78].

**Box 2: (Un)structural methods for unstructured domains: tools for understanding IDR ensembles**

Descriptions of IDR conformational ensembles provide insights into how they carry out numerous and diverse functions in the cell; however, experimental characterization of IDRs is hindered by their flexibility. IDRs generally are not amenable to structure characterization methods like X-ray crystallography and Cryo-Electron Microscopy (Cryo-EM) because their dynamic nature leads to poor electron density. Instead, IDRs are often described by averaged features of the conformational ensemble through solution-state methods including nuclear magnetic resonance (NMR) spectroscopy, small-angle X-ray scattering (SAXS), and Förster resonance energy transfer (FRET) [85–89]. Combinatorially, advances in structure prediction, molecular modeling, and simulations enable access to the structural features of IDRs to shed light on their critical functions [90,91]. Regarding fold prediction with tools like AlphaFold (AF), we urge caution when interpreting the returned 'structures' of IDR-containing proteins. AF predictions notoriously depict IDRs as a sort of 'orange cooked spaghetti' amid a protein's blue globular domains, but recall that IDRs occupy *numerous* conformations.

The structures of the globular domains of the core clock are generally well-characterized. In mouse and *Drosophila* activators, basic helix-loop-helix (bHLH) domains bind DNA. Cryo-EM illuminated how the CLOCK:BMAL1 heterodimer engages nucleosomal DNA to facilitate chromatin remodeling, but the structural model is missing more than half of the protein sequence due to the high flexibility of the IDRs [92]. The PER-ARNT-SIM (PAS) domains in the activators form heterodimers and have been well characterized. *Neurospora* activators have multiple kinds of PAS domains. The LOV/PAS (light, oxygen, voltage) domain is a special subtype of PAS and in WC-1 is important for light entrainment [93] (Fig. 5D). PER/dPER also contain PAS domains, which may homodimerize or heterodimerize with other PER proteins (reviewed in [94]).

CRY1 contains a folded photolyase homology region (PHR) that facilitates its association with PER and CLOCK:BMAL1 in the repressive complex [95,96]. Cryo-EM modeling depicts *Drosophila* TIM using a folded domain and a distant IDR to bind to dCRY, yet the intervening IDRs are absent from the model due to their flexibility [97].

### *IDR-based interactions*

Akin to how an IDR sequence dictates its ensemble properties, sequence features, and ensemble biases govern IDR interactions with other molecules as part of their functions across various cellular processes [79]. IDRs can engage with other biomolecules via conserved sequence features, usually of up to a dozen or so residues, known as short linear motifs (SLiMs) (Fig. 4A) [80]. In these interactions, the bound-state behavior of the IDR falls on a spectrum from disordered to ordered. SLiMs that acquire structure upon binding, often called molecular recognition features (MoRFs), involve a disorder-to-order transition, such as α-helix formation [81–84].

On the other end of the structural spectrum, IDRs can bind in extended conformations wherein the sequence features of the SLiM enable binding to a folded partner [98,99]. Between these extremes lie many dynamic or 'fuzzy' IDR complexes that support a fine-tuned balance of the affinity (strength of the interaction) and specificity (selection of a particular binding partner over another) [100]. Because of their flexibility and solvent exposure, IDRs are poised to interact with multiple and/or distinct binding partners, even using the same SLiM (Fig. 4C) [101]. Finally, like ensemble behavior, IDR binding interactions may be regulated by PTM signals for processes like transcriptional activation and control of protein turnover [102–105].

A key feature of IDR flexibility is the ability to present multiple SLiMs in a single IDR to support one-to-many binding. The ability of a protein to simultaneously interact via multiple distinct binding interfaces is known as multivalency. Multivalent interactions may modulate binding affinity relative to a single SLiM [106–108]. Such interactions can also support higher-order complex assembly by using multiple SLiMs to contact different proteins. Furthermore, these multivalent interactions can lead to the formation of biomolecular condensates [98,109–111] (Fig. 4B). Biomolecular condensates are membraneless compartments that are rich in proteins and/or nucleic acids; these subcellular structures serve to concentrate molecules locally to enhance particular functions [112]. IDRs are common constituents of biomolecular condensates, which are stabilized by many-to-many contacts between molecules. This molecular network can arise from multiple binding motifs and their binding partners [109], or non-specific contacts, like electrostatic or aromatic interactions encoded by the IDR sequence (Fig. 4B) [66,113,114].

### *IDRs in clock proteins*

The flexibility and versatility of IDRs enables them to participate in the most tightly regulated biological processes; therefore, it is unsurprising to find a plethora of IDRs in control of circadian rhythms across species [46]. At the core of clock control, IDRs regulate the activators and repressors alike, in the TTFL (Fig. 5). Moreover, the dynamic nature of IDR ensembles allows them to serve as plastic rheostatic regulators to coordinate timing and integrate cues from the ever-changing environment (Fig. 3A).

Qualitatively, most of the core clock proteins discussed here fall into two categories based on their domain and IDR architecture (Fig. 5). The first category includes those that are enzymes or have homologous roots to enzymes (CK1s/DBT, CRYs, and FRH). All in this category have terminal disordered tails that are important for regulating clock function [29,115,116] (Fig. 5). The remainder of TTFL proteins have an N-terminal dimerization domain(s), a terminal effector or interaction module(s), and a terminal IDR(s) (Fig. 5A). The only exception to these two classifications is CYC, which does not have a terminal IDR (Fig. 5C). An interesting idea to consider is that over evolutionary time the analogous IDR domain architecture of CYC has transferred to other proteins (e.g. CLK, considering its high level of disorder, 72%). A deep sequence space exploration of clock proteins would be an intriguing way to extrapolate on the evolution of IDRs in the clock and the modular nature of domains and IDRs within protein systems.

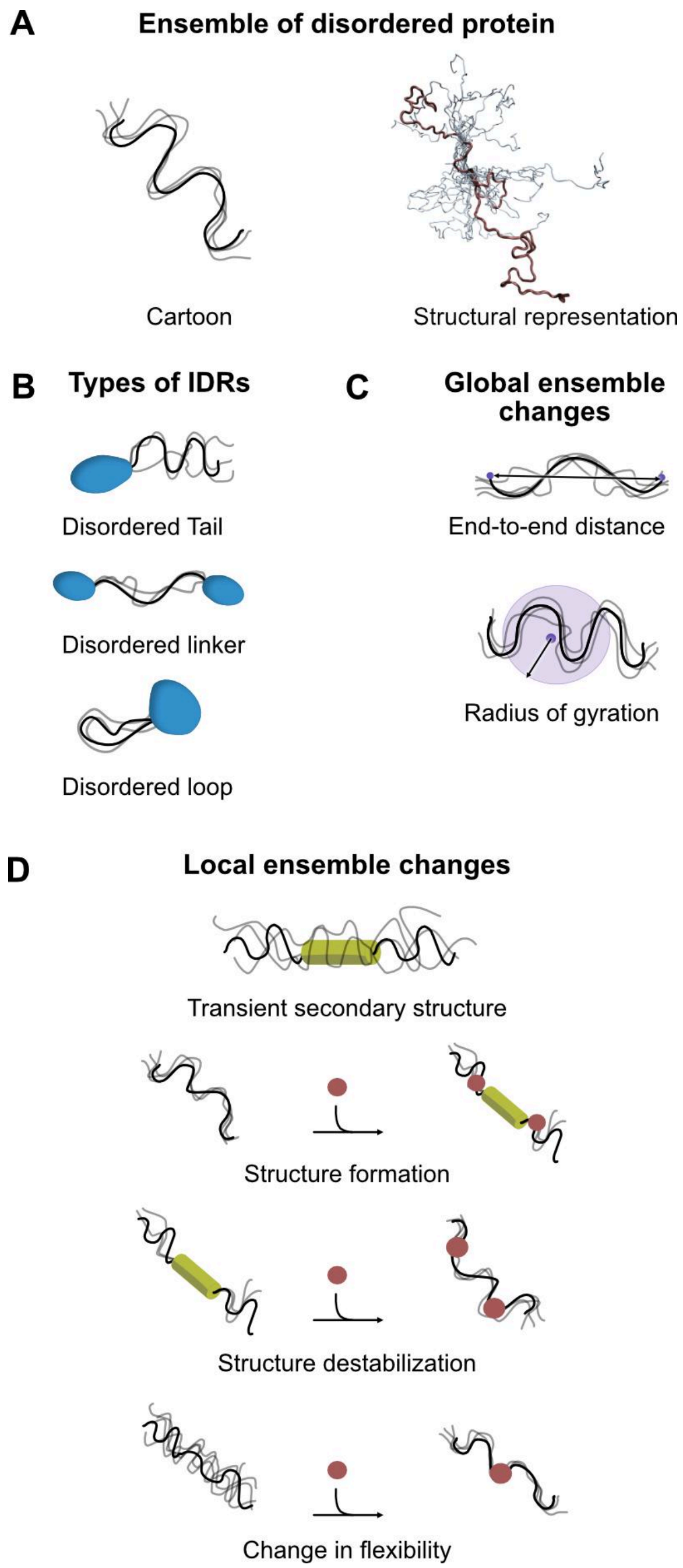

**Figure 2: Ensemble properties of IDPs and IDRs.** IDRs occupy a dynamic, heterogeneous set of interconverting structures known as an ensemble. Depicted in (A) is a representation of a protein that is entirely disordered (IDP). The multiple shaded lines on the cartoon image on the left represent conformations of the ensemble that are energetically accessible, with one conformation darkened in black. To the right is a structural representation of an ensemble from an all-atom simulation of the N-terminus of FRQ, with one conformation highlighted in red. (B) When an IDR is found in the context of structured domains (blue), IDRs are assigned classifications. They are referred to as tails at the terminus, linkers when found between structured domains, and loops when they are excluded from a stable fold within the same protein. (C) Due to the physiochemical properties of the sequence and the environmental context (to be discussed later), IDRs have biases in their ensemble. These biases make it more energetically favorable for particular biophysical features or dimensions of the IDR to be sampled. Some global examples are end-to-end distance, wherein an IDR may be more or less compact. Another example is the radius of gyration. (D) Localized ensemble biases can also result from sequence and context. Transient secondary structure refers to local structural acquisition, for example, transient helicity (yellow). Other examples of local ensemble changes caused by external factors (red circle) can result in a structure formation (often referred to as a disorder-to-order transition), or structure destabilization. Modifications can also change the flexibility of an ensemble.This figure is adapted from Holehouse and Kragelund, 2024 [47].

Enabled by the structure determination techniques discussed in Box 2, the field has made numerous advancements in the biophysical and mechanistic understanding of clock protein structured domains [39,92,97,117,118]. However, due in part to the limited number of tools available to biophysically characterize IDRs in the context of cell biology, the mechanistic understanding of clock protein IDRs has been slow by comparison [47,119]. Over the last few decades, the IDR field has developed approaches and a knowledge base to understand the physicochemical mechanisms underlying IDR function, which has enabled new advancements in applications to clock proteins.

## Biophysical mechanisms of clock protein IDRs

Contrary to early protein structure-function studies, a lack of structure in an IDR doesn't necessarily mean a lack of function. In fact, nearly all biological processes leverage the (un)structural properties of IDRs. Due to their flexibility and sequence features, IDRs are highly versatile, particularly as interaction modules for complex formation. Their extended nature reveals ample surface area for intermolecular interactions and post-translational modifications, both of which are integral to circadian timing.

### *Dynamic macromolecular complex formation and regulation*

IDRs are abundant in transcriptional regulation within transcription factors and their co-regulators, and the clock is no exception [46,47,67,120,121] (Fig. 5). As previously mentioned, the plasticity of ensemble properties and their context-dependence makes IDRs well-suited for progressive processes and the tuning or rewiring of cell regulatory programs. Clock complexes must change dynamically over the day, and IDRs support this, in part, by acting as versatile binding partners or hubs for complex formation. IDRs are naturally suited for multivalency, whether it's through linkers joining structured binding domains or multiple transient interactions with competition for binding and regulation in a context-dependent manner (Fig. 2 and 4).

A key feature of the core clock assemblies is their ability to bind with many partners that regulate function over time. It may seem that IDRs are promiscuous due to their many modes of binding; however, non-specific binding is not a common characteristic, as IDRs can encode specific molecular recognition (reviewed in [47]). As previously discussed, IDRs capitalize on binding specificity through SLiMs (Fig. 4) [122–125]. Some particular examples of clock protein SLiMs include degrons, nuclear localization signals, phospho-sites, and other PTM regulatory modules [45,126–130]. One of the biologically advantageous features of SLiMs is the fact that their ensemble-based mechanisms are often context-dependent, allowing for tuning and large ranges of function (Figs 3 and 4). Core clock protein SLiMs also provide a mechanistic explanation for a portion of the expansive interactome that is centered around the clock complexes [45,130–134].

Within the core oscillator, IDRs enable communication with many binding partners. These interactions may be sequential or simultaneous, and one SLiM may even interact with opposing binding partners depending on the circadian time of day. This is exemplified through the C-terminal IDR of BMAL1, specifically the transactivation domain (TAD). The last 43 residues (comprising the TAD) of BMAL1 are required for transcriptional activation. This same region is required for association with the repressor protein, CRY1 [135].

Both BMAL1 and BMAL2 'TAD' sequences contain a MoRF–transient α-helical bias to their conformational ensembles (Fig. 2D); they differ in that BMAL2 can activate transcription via CBP/p300, but cannot sustain rhythms. Notably, the BMAL2 TAD deviates from the established KIX-binding SLiM, LXXLL [136], but can still bind, albeit with weaker affinity than the BMAL1 TAD. Moreover, swapping the C-terminal IDR in BMAL1 for that of its paralog BMAL2 is insufficient to rescue rhythms in *Bmal1* knockout cells. The evidence that their interactions with CRY1 are similar between the BMAL paralogs but that their motifs fail to rescue one another suggests that the sequence specificity of this IDR is tuned to facilitate a specific interaction with BMAL1 [137].

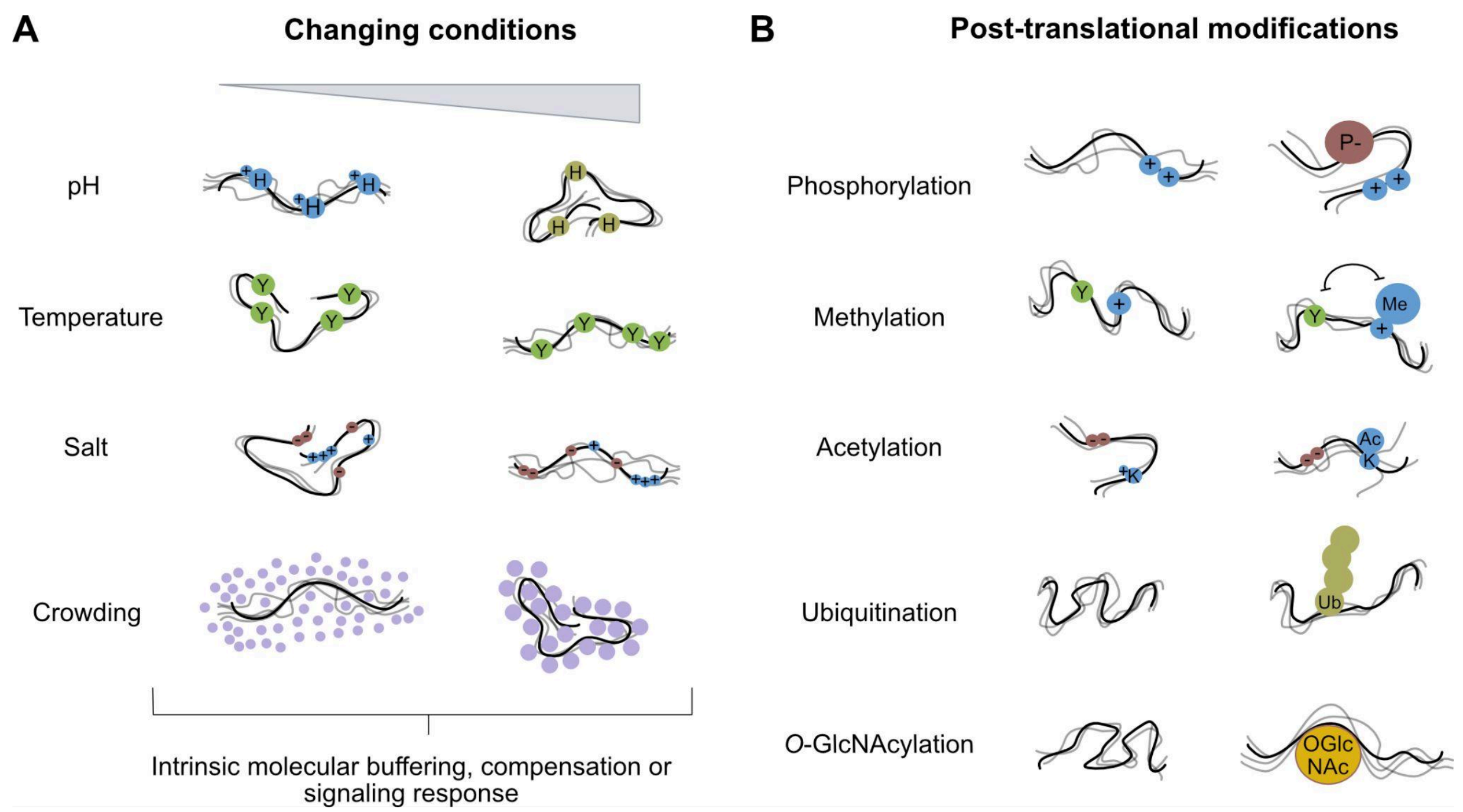

**Figure 3: Physiochemical features that impart IDR-based regulation.** (A) The sequence chemistry of an IDR can be responsive to changes in the solution or environment. pH altering the protonation state of His, Glu, Asp, Lys, and Tyr can result in chain expansion or contraction. Temperature can result in chain expansion from decreased π-π interactions in aromatic residues (shown here, tyrosine). Salt can modulate electrostatic interactions to alter ensemble dimensions in a sequence-dependent manner. Depending on the shape, chemistry and size of crowders, ensemble dimensions can also vary. (B) Biological regulation can impart changes in ensemble behavior and dimensions via post-translational modifications (PTMs). Phosphorylation can cause chain compaction or alter long-range intramolecular interactions based on IDR charge or sequence features. Methylation of arginine results in reduction of cation-π interactions with aromatic groups like tyrosine leading to IDR dimension changes. Acetylation results in the loss of the positive charge of lysine, which could influence ensemble properties based on sequence features (e.g., a proximal negative charge). Large modifications like glycosylation or the addition of a polypeptide (ubiquitination or SUMOylation) can alter the accessible ensemble space due to steric constraints. These modifications may also drive other ensemble changes based on the IDRs sequence or prevent binding and access of other macromolecules. This figure is adapted from Holehouse and Kragelund, 2024 [47].

Another repressor protein, CHRONO, interacts with the BMAL1 TAD with higher affinity than either the CBP/p300 KIX domain or CRY1. CHRONO binding is specific to the BMAL1 TAD, and it does not interact with BMAL2 [138]. The binding promiscuity of the BMAL1 TAD to both activator and repressor proteins illuminates the advantages and regulatory capabilities granted by IDR ensembles and SLiMs; yet, the specific signals and mechanisms that distinguish between binding partners remain unclear.

A co-crystal structure of CRY1-PER2 highlights the advantage of solvent-exposed features in IDRs; the ~80-residue CRY-binding domain (CBD) of PER2 is well-resolved in this structure and binds the PHR surface in a largely extended conformation. This binding mode serves to modulate CRY1 serine loop dynamics to tune CRY1 affinity for CLOCK:BMAL1 [95]. Other structures of CRY-PER paralogs depict a similar binding mode for PER–extended across the CRY surface–and implicate a zinc coordination site formed at the interface of the two molecules [139,140].

Multivalent interactions are common features of IDR-containing complexes, such as those found in the CRY1-PER2-CLOCK:BMAL1 complex. The repressor complex consisting of CRY1, PER2, and CLOCK:BMAL1 is integral to appropriate timekeeping, wherein CRY1 PHR interacts with CLOCK PAS-B (via the 'secondary pocket'). In addition, IDRs of PER2 (across the PHR surface), and BMAL1 TAD (via the CC helix) contribute to this complex [139,141] (Fig. 5B). Curiously, the TAD is more than 100 residues away from the PAS-B domain (Fig. 5B), which underscores how IDR flexibility can contribute to the 'reach' of SLiMs during complex formation.

The power of long-range interactions facilitated IDR flexibility is exemplified by the repressors FRQ/PER/dPER. In *Neurospora*, CK1a associates with FRQ through a bipartite interaction between the kinase and two partially helical SLiMs in FRQ. Notably, these SLiMs (FRQ-CK1 interaction domains, FCD1 & FCD2) are more than 150 residues apart in linear sequence space (Fig. 5D). Deletion of one of these SLiMs renders CK1a incapable of stable association and phosphorylation of FRQ; however, relocating one FCD to the FRQ N-terminus permits this functional interaction [142]. These findings suggest a model in which CK1a binding leverages FRQ flexibility to bring the FCD SLiMs together in 3D space, despite their separation in sequence.

CK1/DBT interactions in PER/dPER similarly require a >100-residue disordered region for binding and target phosphorylation [143][26]; inspecting the disorder predictions for all three repressor homologs reveals two short motifs of less disorder (lighter color) within the longer CKBD region (Fig. 5C-D). The stable kinase interaction with two distant SLiMs appears to be a conserved feature of the repressive complex. In a similar vein, TIM interacts with *Drosophila* CRY–which serves a photosensing role–via two separate elements. A Cryo-EM structure revealed first an interface between folded domains formed by the N-terminal Armadillo (ARM) 1 domain of TIM; the second motif falls within the C-terminal predicted IDR (>1000 residues away), and EM modeling suggests this is a folding-upon-binding interaction [97].

In line with the strict regulatory requirements that support timekeeping, IDRs not only facilitate intermolecular interactions, but also interactions in *cis* through IDR contacts with a folded domain on the same protein chain (intramolecular interactions). One such example is the competition of the CRY1 C-terminal IDR ('tail') with the CLOCK PAS-B domain for binding to the CRY photolyase homology region (PHR) (Fig. 5B). Deletion of CRY1 exon 11 of the tail modulates the affinity of PHR to CLOCK:BMAL1 and lengthens the period in cells by enhancing repression. Notably, the CRY1 tail fails to compete with PER2 for binding, given a nearly 100-fold difference in the binding affinities for PHR between the two partners [144].

Another example of 'auto-regulation' of a clock protein via its IDRs is the CK1 autoinhibitory tail, discussed in more detail below, in which CK1 activity on its own C-terminal IDR prompts interaction between the phosphorylated IDR and the catalytic domain to inhibit function toward PER2 [115,145]. Outside of the core clock components, a point mutation in the C-terminal predicted IDR of salt-inducible kinase 3 (SIK3) is associated with a 'natural short sleep' phenotype in mice. Curiously, this point mutation decreased kinase activity in *in vitro* assays, suggesting that the mutation has a direct impact on catalytic function despite being hundreds of residues downstream of the active site [146]. Given the evidence in other clock proteins for IDR-driven self-interactions, we could envision a model in which long-range interactions of the SIK3 IDR with the globular kinase domain modulates activity toward SIK3 substrates.

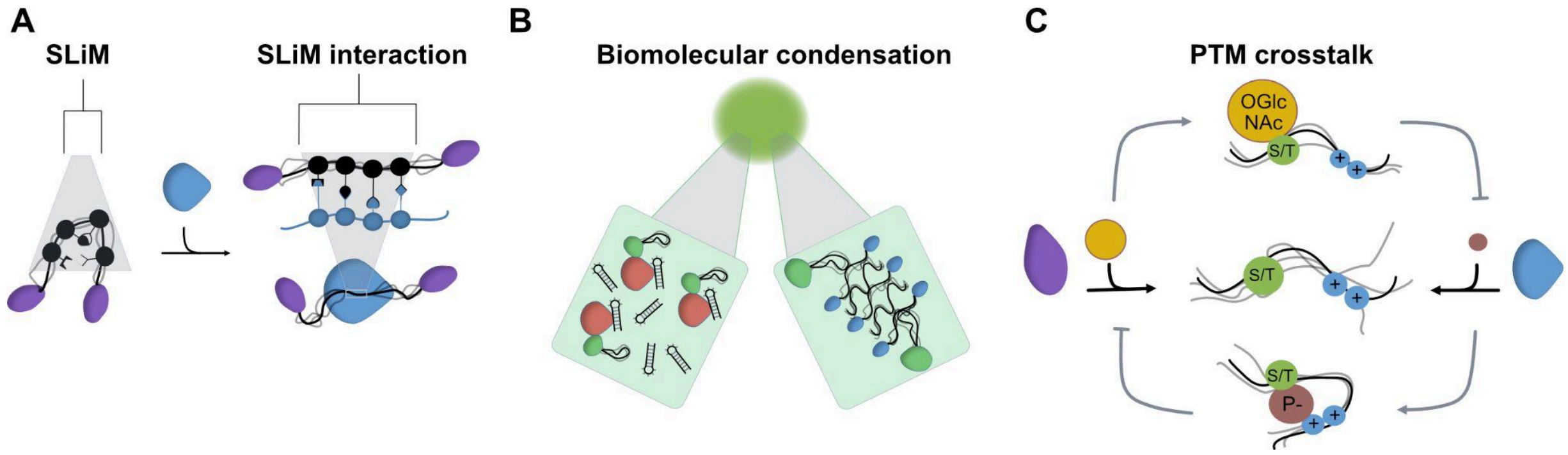

**Figure 4: IDR interactions and emergent processes.** (A) Short Linear Motifs (SLiMs) serve as interaction modules within disordered regions. The specific shapes within the interaction region cartoon represent complementary chemistry, which facilitates binding between two constituents. (B) The properties of certain IDRs in conjunction with other components (i.e., RNA or other binding partners) can facilitate condensate formation. IDRs in these contexts can serve as binding regulators or scaffolds that enable the proper intermolecular architecture, yielding energetically favorable conditions for condensate formation to occur. (C) PTMs can influence one another. Depicted here is the example of phosphorylation and *O*-GlcNAcylation of the same Ser/Thr. Blue is a kinase and purple is a *O*-GlcNAc transferase. In specific biological contexts, one modification could block the other, preventing/facilitating downstream processes (e.g., degradation, phase separation, etc.). Considering that both modifications are reversible, it serves as an excellent platform for the fine-tuning of regulatory processes.

IDRs mediating repressive complex formation are abundant in the *Neurospora* clock as well. FRQ is stabilized by FRH, an RNA helicase and homolog of MTR4–a coenzyme for the nuclear exosome that facilitates polyadenylation [117,147]. While the helicase function of FRH is not necessary for clock function, the intrinsically disordered N-terminus of FRH (150 aa) is necessary for interaction with FRQ and, therefore, clock function [147]. The current model is that FRH serves as a ‘nanny’ by immediately binding FRQ upon its translation to stabilize and prevent its premature degradation.

Another modulator of inter- and intramolecular interactions in IDRs is sequence charge patterning (Fig. 2B and C) [67,148–151]. Clock repressor proteins in *Neurospora, Drosophila,* and mammals are enriched in positive charge blocks in their IDRs, suggesting that distributed positive charge plays a role in complex regulation. Moreover, these positive charge blocks have been shown to impact output and period length [152]. Considering the conservation of positive charge blocks in clock repressors and their extensive progressive phosphorylation over circadian time suggests a model where disorder-mediated charge dynamics are critical for proper complex formation and clock regulation.

### *Post-translational modifications and their effects on the core clock*

#### Phosphorylation

Post-translational modifications (PTMs)–particularly phosphorylation–are integral to the TTFL. Clock speed is choreographed by a cycle of complex formation, phosphorylation, and degradation of the core repressor proteins (PER/dPER/FRQ) [153]. Consequently, dysregulation of the repressing kinases (CK1 in mammals, DBT in *Drosophila*, and CK1a in *Neurospora*;

collectively called ‘CK1’ here) or their phosphorylation targets leads to incorrect timekeeping [37,154,155]. CK1 targets within the core clock include both activators and repressors, but the common theme is that the overwhelming majority of modifications tend to occur in IDRs [45]. Including the CK1 sites, nearly all of the 60 putative PER2 phosphosites fall in predicted IDRs, which comprise 61% of the PER2 structure (1257 residues) [27,156,157] (Fig. 5B). Similarly, dPER (1224 residues, 72.3% disordered) contains at least 64 phosphosites and FRQ (989 residues, 74.4% disordered) harbors more than 100, with the majority of sites falling within their IDRs [27,45,129,130,158] (Fig. 5C-D).

CK1 associates with designated CK1-binding domains on the repressors (CKBD in PER2 and dPER, FCDs in FRQ) to control their stability and, hence, period length (Fig. 5) [159]. Phosphorylation motifs in PER2 are also SLiMs and include the FASP region and phospho-degrons [160]. The familial advanced sleep phase (FASP) region falls in the CK1-binding domain (CKBD, Fig. 5). It is named for the shortened period arising from a serine to glycine mutation that alters the motif sought by CK1 and redirects the kinase toward the degron motif to drive PER2 turnover [37]. In the wild-type case, sequential phosphorylation of the FASP region leads to CK1 product inhibition, decreased activity toward the phospho-degron, and ultimately increased PER2 stability; this is collectively called the PER2 ‘phosphoswitch’ [161]. Phosphorylation at the degron site(s) in these proteins creates a SLiM recognized by a ubiquitin ligase (β-TrCP in mouse, SLIMB in *Drosophila*, and FWD1 in *Neurospora*) that facilitates the proteasomal degradation of the clock repressors according to the time of day [28,127,159,160,162,163]. CRY2 is also proposed to be a target of β-TrCP via phosphorylation of its C-terminal IDR by GSK-3β [164].

FRQ has two reported ‘PEST’ sequences, which are rich in the disorder-promoting residues proline (P), glutamate (E), serine (S), and threonine (T) (Fig. 5D). A PEST motif is a type of SLiM often involved in protein turnover mechanisms, which is usually prompted by a phosphorylation event. Given their sequences and role as interaction motifs, PEST sequences are more frequently found in IDRs, where they are readily accessible to interactors compared to folded domains [165,166]. In FRQ, both PEST sequences fall immediately C-terminal to a folded domain and at least one works in concert with a nearby phosphosite to control FRQ stability; mutation of this or nearby phosphosites in the first PEST region leads to FRQ hypophosphorylation, decreased protein turnover, and, hence, a longer period [28,128,130]. Mutation of C-terminal FRQ phosphosites (serine to alanine) has profound but opposite implications for FRQ stability and period length, which reflects the versatility of phosphorylation signals [130].

Similarities across species in phosphorylation-controlled repressor protein stability highlight the versatile and critical regulatory roles carried out by IDRs. Given the prevalence of disorder and extensive phosphorylation in the repressor proteins beyond the two aforementioned motifs, models have emerged that ascribe a global role for hyperphosphorylated PER2/dPER/FRQ. The CK1-binding domain SLiMs on PERs and FRQ are indispensable for the phosphorylation of the many sites that lack a priming phosphorylation mark for CK1. The stable association of CK1 with the repressor enables its phosphorylation of these non-canonical motifs, while highly flexible IDRs permit access to the kinase active site, even from hundreds of residues away [29,143,167]. Biophysical measurements of FRQ complexes with and without hyperphosphorylation point to a model of repressor ensemble expansion and complex dissociation as a result of CK1 (and other kinase) actions [45,142,168,169]. Collectively, the variable and tunable phosphorylation events in PER2/dPER/FRQ IDRs confer a crucial TTFL timing delay under the control of CK1 and its isoforms [161,170–172].

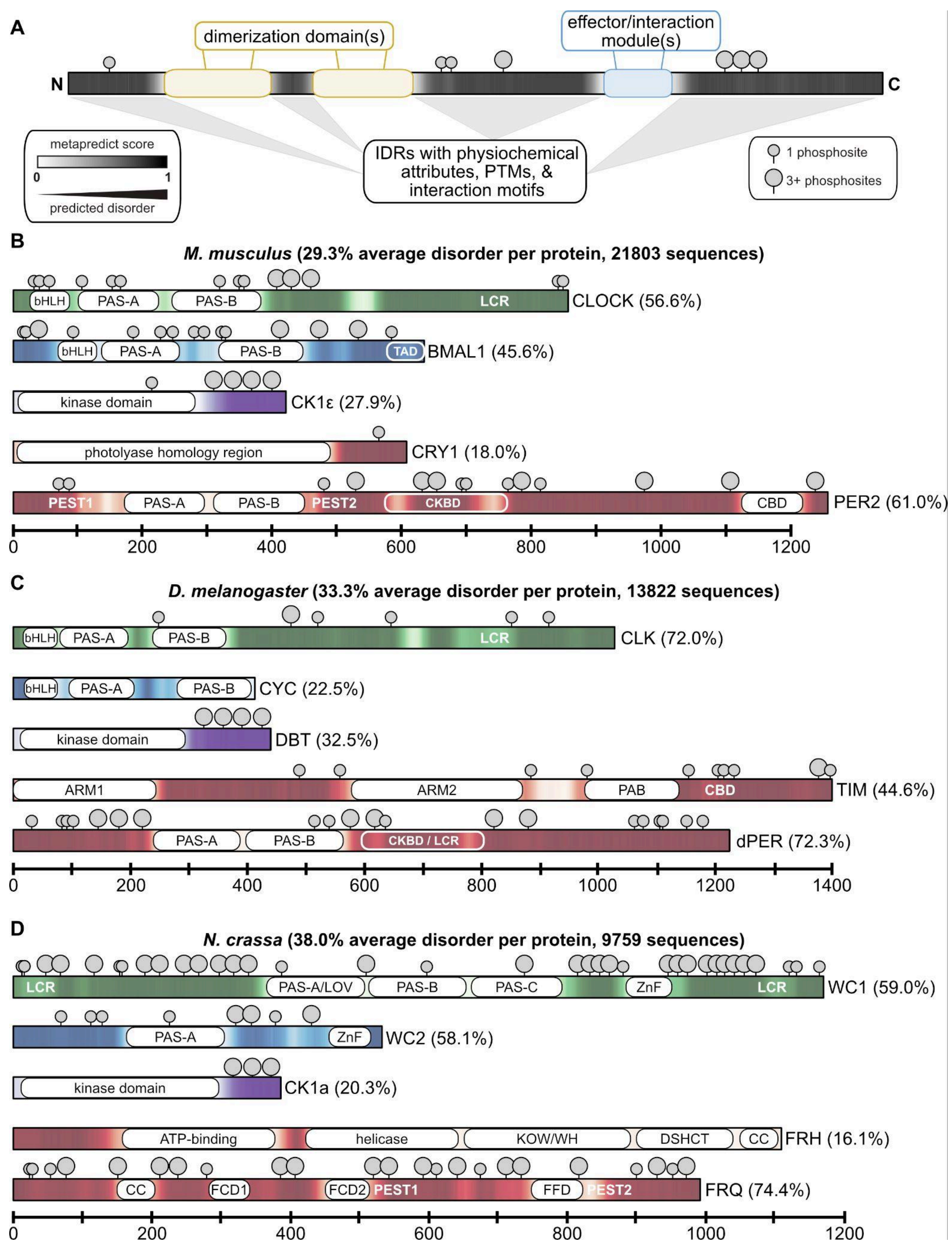

**Figure 5: Clock proteins across species share features and disorder propensities.** (A)

Representative domain diagram of a core clock protein containing folded dimerization (yellow) and effector binding (blue) domains amidst an otherwise disordered protein (grey to black). Shading corresponds to the per-residue disorder score (0, least disordered, to 1, most disordered) calculated with 'metapredict' [55]. Protein regions are colored such that the darkest coloring represents high predicted disorder and white reflects predicted folded regions. Putative (non-exhaustive) phosphorylation sites are shown with grey circles. (B) Domain diagrams for the mouse core clock proteins shaded by disorder score. Structural and/or functional domains are labeled in each diagram. Percent disorder for each protein is given in parentheses. (C) Domain diagrams for the *D. melanogaster* core clock proteins, colored the same as in (A). (D) Domain diagrams for the *N. crassa* core clock proteins, colored by disorder score according to the color bar in (A). For all three species, reference proteomes containing one protein sequence per gene were downloaded as FASTA files from Uniprot (July 2025). Abbreviations can be found in list form in the abbreviations box.

Other repressor proteins, CRYs (mammals) and TIM (*Drosophila*), also harbor phosphorylation sites in their IDRs (Fig. 5). CRY1 can be phosphorylated on its C-terminal IDR to block binding by a ubiquitin ligase complex. A constitutively-phosphorylated mutant of CRY1 (serine to aspartate phosphomimetic) has a longer half-life and lengthens the period [173], highlighting the connection between IDR phosphorylation and period maintenance. Similarly, deletion of a serine-rich portion of the TIM linker IDR containing putative CK2 and DBT phospho-sites increased the behavioral period, suggesting a possible role for phosphorylation in TIM function control [174]. These phosphorylation sites were assigned a role in the nuclear import of the dPER/TIM complex, catalyzed by the kinases SHAGGY and CK2 [175,176].

Compared to other core clock proteins, CK1 has relatively low disorder (Fig. 5B-D). However, each CK1/DBT/CK1a harbors a C-terminal intrinsically disordered tail that is a critical determinant of its catalytic activity. In addition to an auto-phosphorylation site in the CK1 catalytic domain that alters substrate specificity [177], there are also ten additional reported sites of auto-phosphorylation throughout the C-terminal IDR that contribute to enzymatic regulation of the human CK1δ isoform [178]. Consistent with this model, early studies of CK1δ and CK1ε activity noted that truncation of the IDR or treatment with a phosphatase augmented kinase activity [145,179]. *In vitro* studies of CK1δ, CK1ε, and the *S. pombe* homolog Hhp1 established that the phosphorylated tail is capable of a weak, fuzzy interaction with the catalytic domain in *cis* to inhibit activity [180]. In particular, as determined for the CK1δ1 isoform, one of three unique phospho-sites may occupy one of the anion-binding sites on the kinase domain to block substrate recognition; mutation of these three phospho-sites yields a shorter period (~3 h) in luciferase reporter assays [115]. In this example, the auto-phosphorylation behavior of CK1 induces long-range intramolecular interactions by a C-terminal IDR to modulate catalytic output and, therefore, timekeeping.

Inactivating autophosphorylation of CK1 in cells is opposed by the actions of phosphatases, which support a cycle of kinase activation and inactivation through dynamic PTMs [181]. Phosphatase activity toward PER2 likewise opposes the actions of CK1δ, such that the phospho-state of PER2–a key determinant of oscillator speed–is tightly controlled [182]. Additionally, PER2 association with CK1ε is reported to inhibit its autocatalytic function and increase activity toward CRY2; rhythmic nuclear import of PER2 exerts temporal control over its role as a scaffold for recruitment of CRY2 to the PER2-CK1 complex [183,184]. The intricacies of the repressive complex highlight the importance of IDRs in higher-order complex assembly dependent on their PTM status and multivalent interactions with other molecules.

Phosphorylation of clock protein IDRs is not limited to the repressive arm of the TTFL. The *Neurospora* activators, WC-1 and WC-2 (together, WCC), can undergo 15 and 113 putative phosphorylation events, respectively, the vast majority of which are within their IDRs (Fig. 5D). Notably, several phospho-sites in WC-1 lie near the ZnF domain, which could reflect a model by which IDR phosphorylation near a DNA-binding domain prevents association with DNA via electrostatic repulsion [185]. Phosphorylation of the White Collar Complex is also FRQ-dependent [186], which underscores the functionality of flexible IDRs that can recruit and sustain interactions with effector proteins, such as CK1. In a similar way for mammals, treatment of CLOCK:BMAL1 complexes with CK1δ reduced binding of the complex to DNA [187], whereas treatment with the CK1ε isoform appeared to *increase* transcriptional activity of CLOCK:BMAL1 [184]. In both cases, phosphorylation of CLOCK:BMAL1 and the consequent regulatory outcomes are dependent on CRY and PER [188,189], which demonstrates the utility of dynamic domain interactions across complexes with different compositions and roles.

### Methylation and Acetylation

Lysine and arginine methylation (the addition of a methylene group to a lysine Nε or arginine guanidinium moiety) and lysine acetylation (the addition of an acetyl group to a lysine Nε), like phosphorylation, are reversible post-translational modifications that modulate the sequence features of a protein toward a particular interaction or function. Methylation and acetylation are heavily studied in histone tails, where they demarcate 'active' and 'repressed' genes through selective recruitment of chromatin remodelers and transcription factors. However, these modifications still occur in many other proteins [190]. Mass spectrometry analysis of WC-1 and WC-2 from cultures grown in constant light mapped methylation and acetylation sites across both proteins. All six WC-1 acetylation sites and two of three sites in WC-2 lie in folded domains. Wang and colleagues identified nine methylation sites in WC-1; however, only two sites fall in IDRs. Site-specific mutations to block acetylation or methylation had little effect on rhythms in *Neurospora* [191], which suggests that activator methylation and acetylation may play a role outside the core timekeeping mechanism.

BMAL1 acetylation has been reported in its C-terminal IDR, which includes its transcriptional activation domain (TAD) at the distal C-terminus [135]. However, the acetyllysine marks do not fall within the TAD itself. One site, which was identified by a large-scale acetylome dataset as a CBP/p300 target, is about 100 residues upstream of the TAD [192]. The other site, a TIP60 target, lies about 50 residues upstream of the TAD and appears to be rhythmically acetylated at the onset of repression [193,194]. Notably, before the identification of TIP60, CLOCK was thought to be the acetyltransferase responsible for BMAL1 modification [193,195]; however, it is unlikely that CLOCK could perform this action given its high degree of disorder outside the dimerization and DNA-binding domains and its lack of a catalytic domain (Fig. 5B).

Separate studies report different potential functions for the TIP60 acetyllysine mark: one found that acetylated BMAL1 recruits the transcription elongation factor, P-TEFb, via its bromodomain [194]. Another study found that acetylation promotes BMAL1 association with CRY [193], potentially by altering the conformational ensemble of the IDR via charge neutralization [196]. Although the specific consequences of acetylation and prevalence in other clock proteins are still unclear, these examples demonstrate how PTMs can tune IDR conformations or modify SLiMs toward specific regulatory outcomes.

### *O*-GlcNAcylation

*O*-GlcNAcylation is a reversible PTM that covalently attaches an *O*-linked *N*-acetylglucosamine (*O*-GlcNAc) group to serine and threonine residues. *O*-GlcNAcylation plays a role in diverse biological processes, including nutrient and stress response regulation [197]. Mechanistically, *O*-GlcNAcylation has been identified to regulate circadian repression and feeding rhythms [198]. Like phosphorylation, *O*-GlcNAcylation primarily occurs within disordered regions [74,199,200].

*O*-GlcNAc moieties have been identified on several clock proteins, including CLK, dPER, and PER2 [198,201].

At the molecular level, O-GlcNAcylation of dPER Ser942 occurs within its disordered C-terminus and CLK binding region (Fig. 5C). dPER S942A increases the interaction between dPER and CLK, suggesting that O-GlcNAcylation decreases the affinity of the dPER C-terminus for CLK. Such regulation could arise by changes to the local ensemble or through alteration of interaction sites via the *O*-GlcNAc modification(Fig. 3B). Notably, the dPER binding region in CLK is in its C-terminal IDR (Fig. 5C), suggesting that *O*-GlcNAc tunes an IDR-IDR interaction between CLK and dPER. Moreover, the S942A mutation causes alteration in feeding times and a short period phenotype, supporting a model where *O*-GlcNAcylation of S942 prevents the early start of the repression phase [198].

In a recent report, Liu and colleagues surveyed the liver proteome over time, specifically investigating changes in *O*-GlcNAcylation and phosphorylation [202]. This dataset revealed many modified substrates with considerable crosstalk between *O*-GlcNAcylation and phosphorylation (Fig. 4C). Their work identified several proteins that were rhythmically phosphorylated and glycosylated, with a correlation between the two, suggesting the two PTMs may regulate each other. Notably, they found that *O*-GlcNAcylation and phosphorylation of CLOCK are nearly antiphase with one another [202].

At the site-specific level, they investigated CLOCK S431, which was identified as both phosphorylated and *O*-GlcNAcylated. Their findings suggest that S431 *O*-GlcNAcylation competes with S431 phosphorylation, in turn regulating S427 phosphorylation to modulate CLOCK transcriptional activity. The authors show that altered *O*-GlcNAcylation of this region enhances repressor activity, which parallels previous work demonstrating mPER2-S662G (*M. musculus*) results in an enhanced repressor activity relative to WT [204]. These findings suggest that temporally opposing PTMs drive molecular behavior to facilitate S427 phosphorylation and CLOCK degradation at the correct time of day. This model positions PTM crosstalk in IDRs as an important mechanism for regulating repressor activity and period length.

### Ubiquitination and SUMOylation

Ubiquitination and SUMOylation are reversible PTMs with roles in protein stability and turnover, as well as DNA damage response and signal transduction [205]. Unlike the other modifications discussed here, ubiquitination and SUMOylation involve the covalent attachment of a protein molecule (ubiquitin, Ub, or the small ubiquitin-like modifier, SUMO) to a lysine side chain (Fig. 3B). Both modifications are installed via similar enzyme cascade pathways, are reversible via Ub/SUMO-specific proteases, and serve critical roles across eukaryotes, including within the core clock [206,207].

One common function of ubiquitination is to target proteins for degradation. Poly-ubiquitination and proteasomal degradation of the PER proteins, prompted by their phosphorylation, is a crucial mechanism for de-repression of the activators [159,160]; this process is mirrored in *Drosophila* and *Neurospora*, wherein multi-site phosphorylation drives repressor protein turnover via SLIMB and FWD1, respectively [162,163]. Once phosphorylated, PER2 associates with the ubiquitin ligase β-TrCP via two degrons within the PER2 IDRs. The installation of the phosphate creates a SLiM compatible with β-TrCP binding to facilitate ubiquitination on a handful of lysine residues in the folded and disordered domains of PER2 [208]. CRY levels are also rhythmically regulated by a ubiquitin ligase via its association with the folded PHR domain [209].

The ubiquitin ligase UBR5 was recently identified to modify BMAL1. Accordingly, siRNA depletion of this enzyme in cells led to increased BMAL1 stability and transcriptional activity, as well as a shorter period; the *Drosophila* activator, CYC, is downregulated similarly by its UBR5 homolog [210]. Before this, SUMOylation within the intrinsically disordered linker between PAS domains was established as an important determinant of BMAL1 stability. Mutation of the putative SUMOylated lysine led to increased BMAL1 stability, suggesting crosstalk between the SUMO modification and the BMAL1 degradative pathway [211]. A later study reported the localization of SUMOylated BMAL1 to PML (Promyelocytic Leukemia) bodies and that ubiquitination occurred within these compartments [212].

There is still an outstanding need for more finely resolved spatiotemporal information about clock protein modifications and their regulatory/phenotypic outcomes. At the heart of physiology and adaptation, the core clock represents an opportunity to observe the dynamic and versatile behaviors of IDRs and how they can regulate–or be regulated by–their environments through post-translational modifications.

## Disorder-associated emergent phenomena: mechanisms in circadian physiological regulation and molecular entrainment

### *IDRs in signal processing and propagation*

The cyclical environment has periodic temperature variation, which can signal and influence clock timing. Consequently, temperature can entrain or synchronize rhythms in a population of cells or organisms [213–215]. Because chemical and biological reaction rates are sensitive to temperature changes, clocks have also evolved buffering mechanisms to differentiate temperature noise from signal. The phenomenon of buffering rhythms against environmental fluctuations to maintain a homeostatic clock period is referred to as temperature compensation (reviewed in [216]). While some compensation and entrainment parameters have been characterized at the molecular level, there is still much to learn about the intersection of temperature compensation and the temperature entrainment response.

One mechanism associated with the temperature response in eukaryotes is temperature-induced alternative splicing that yields differential proteoforms of the core clock proteins [217–221]. Alternative splicing in the clock often influences the presence or length of an IDR [115,217,218,222]. In this way, splicing can alter protein interactomes, rewire transcriptional regimes, alter protein stability, signaling, and spatiotemporal regulation [223–229]. This suggests that the clock uses spliced IDR variants to serve as tunable protein modules to support robust timekeeping.

In *Neurospora*, *frq* is alternatively spliced in a temperature-dependent manner, such that the difference between the two isoforms is an N-terminal IDR of 99 amino acids [218–220]. The clock can function in a mono-isoform manner, but maintaining robust rhythms across a range of temperatures requires both isoforms [220]. At elevated temperatures, the isoform with the lengthened N-terminal IDR is the dominant isotype. Yet, the mechanistic advantage conferred by this IDR and its relation to temperature response and compensation remain unknown.

While temperature compensation is clearly crucial for ectotherms and microbes, it is also important for endotherms [230]. In mammals, the current model for temperature compensation posits that differential phosphorylation of two regions in PER2 affects its stability. Importantly, temperature has been shown to direct the substrate selection of mammalian CK1, which influences repressor protein stability [231,232]. Additionally, PER2 phospho-priming varies between CK1 isoforms, and phosphokinetics are slower on unprimed sites [232].

*Drosophila* temperature compensation has primarily been ascribed to phosphoclusters within IDRs that regulate the stability of dPER in a temperature-dependent manner [162,216,233]. However, limit-cycle modeling suggests that TIM protein level oscillations influence temperature compensation, which fits the repressor stability model of compensation, as TIM regulates dPER [216,234,235]. Temperature-induced alternative splicing in *Drosophila* produces multiple TIM proteoforms with differing IDRs [236]. One splice variant removes a section of the first IDR toward the N-terminus, whereas other variants affect the integrity of the PAB domain and C-terminal IDR (Fig. 5C). In support of a role for variable TIM IDRs in temperature/seasonal adaptation in *Drosophila*, mutations in these IDRs alter temperature compensation in vivo [237].

Repetitive IDR sequences, known as LCRs, have been found to serve temperature-sensing roles in plants and fungi [61,238–240]. In the *Arabidopsis thaliana* (At) clock, LCR poly-glutamine (polyQ) stretches have been shown to play a role in modulating circadian output in a temperature-dependent manner [239]. Furthermore, the length of the polyQ repeats and other LCRs in *Drosophila*, *Neurospora*, and vertebrate clock proteins have latitude-dependent length variation[241–249] (Fig. 5), suggesting that evolution has tuned LCRs in IDRs to serve as rheostats for temperature and seasonal adaptation.

In plant clocks, CRY is a light sensor and a source of input into the clock. One of the very first examinations of disorder in circadian clock proteins centered around the disordered C-termini of Cryptochrome proteins in *Arabidopsis* and animals [250]. This work demonstrated that the disordered C-terminal IDRs of AtCry1 and hCRY2 interact with their respective photolyase-homology regions. The authors provide further evidence that this is a light-dependent phenomenon in AtCry. Specifically, the authors found that light could induce an order-to-disorder change, which was later validated using orthogonal methods by Kondoh and colleagues [251]. The examples here of temperature- and light-responsive IDRs illuminate a trend in clock protein IDRs to sense and respond to external cues.

### *Biomolecular condensates around the clock*

Biomolecular condensates are membraneless compartments that recruit certain biomolecules while excluding others [112,252–254]. IDRs are known to regulate or contribute to the formation of biomolecular condensates, although not all IDRs are condensate-associated, and structured domains can also serve as condensate drivers and regulators [66,122,255–258]. Biomolecular condensates play many roles in the cell, regulating processes such as stress regulation, nucleic acid processing, DNA damage repair, and transcription [112,257,259,260]. It follows that IDR-mediated biomolecular condensate formation could serve as a mechanism to enhance the biological timing of many processes both within and outside the core clock.

In the *Neurospora* core clock, evidence suggests that condensates regulate timing at multiple levels [46,168,261]. At the post-transcriptional level, the *frq* and *ck1a* transcripts are regulated by the largely disordered RNA-binding protein PRD-2. Bartholomai and colleagues demonstrated that PRD-2 forms dynamic puncta with seemingly coordinated movement *in vivo* and condensates of higher-order structure *in vitro*, both of which contain *frq* [261]. Furthermore, PRD-2 protects *ck1a* from nonsense-mediated decay, and knockout of PRD-2 results in a long-period phenotype (+3.5 hrs) [262]. Considering these data together, this suggests a post-transcriptional regulation mechanism in which PRD-2 employs condensates to support the robust spatiotemporal coordination of key clock protein transcripts. Furthermore, as IDR ensembles and biomolecular condensates are tunable by environmental factors such as temperature (Fig. 2) [239,263,264], PRD-2 condensates could serve as a regulatory rheostat for temperature response or compensation by controlling the processing time of *frq* and *ck1a*.

Post-translationally, FRQ protein has been shown to form biomolecular condensates *in vitro* and puncta *in vivo* [168]. Tariq and colleagues employed several in vitro biophysical methods to characterize the stoichiometry of FRQ, its relation to CK1, including phosphokinetics, and the crosstalk with phase separation. They found that both unphosphorylated and phosphorylated FRQ exhibit foci formation, although their experiments were performed at different salt concentrations, which is known to influence condensate formation. They demonstrated that FRH and CK1 are recruited into FRQ droplets. Intriguingly, they showed that FRQ phosphorylation via CK1 is reduced in droplets, which could reflect an emergent regulatory mechanism or a byproduct of altered enzyme kinetics in the puncta [168].

Another example of IDR-driven condensate regulation of circadian physiology stems from REV-ERB$\alpha$, an auxiliary repressor protein that regulates the expression of BMAL1 and various *ccgs.* Recent work has shown that a central ~200 amino acid IDR (referred to as the hinge IDR, or hIDR) regulates rhythmic gene expression via condensate formation. The authors demonstrate that the hIDR is responsible for repressive transcriptional hub formation in the liver. Their evidence indicates IDR-IDR interactions selectively partition co-repressor NCOR1 into nuclear condensates [265]. NCOR1, a 2,453 amino acid IDP (~90% disordered), is a known transcriptional regulator that facilitates gene repression by histone acetylation regulation [266,267]. Their model suggests that REV-ERB$\alpha$ facilitates repression by timing the formation of chromatin hubs that recruit NCOR1 and co-regulators, a process that is orchestrated by biomolecular condensates (Fig. 4B).

The *Drosophila* molecular oscillator has also been shown to use biomolecular condensates as a means of spatiotemporal organization of transcriptional activity [268]. CLK and dPER foci have been identified at the nuclear periphery during critical stages of repression. The foci were shown to play an important role in the subnuclear localization of core clock genes. The current model posits that clock protein foci coordinate temporal clustering and reorganization of clock-regulated genes during the repression phase. Notably, CLK and dPER have LCR sequences in their IDRs. Previous work has identified a LCR domain in a plant clock protein that drives phase separation and regulates stress responses [238,239]. This supports the hypothesis that LCRs may contribute to dPER and CLK foci formation [46].

## Conclusions and perspectives

IDRs play a crucial role in circadian timekeeping across the eukaryotic domain. Despite their lack of a stable 3D structure, IDRs serve as versatile and dynamic interaction modules within the core TTFL. The plasticity and dynamic nature of IDRs enable them to regulate both activation and repression mechanisms that are essential for circadian precision and robustness. Furthermore, IDRs are also poised to facilitate circadian regulation outside of the core TTFL mechanism, broadening the reach of the clock into many aspects of physiology. The conservation of IDRs in clock proteins from fungi to mammals underscores their fundamental role in circadian biology. Understanding the contributions of these disordered regions not only sheds light on the molecular architecture of the clock but also broadens our understanding of how disorder facilitates regulation in complex cellular systems. The mechanisms and molecular details discussed here are only a fraction of the immense body of work in the circadian field to date, and we regret the work we were not able to discuss due to the constraints of space and time.

## Acknowledgements

We thank Drs. Carrie Partch and Alex Holehouse for their feedback on this manuscript. Uniprot IDs used to create Fig. 5: Q9WTL8-1, O08785-1, Q9JMK2, P97784, O54943, O61735-1, O61734, O76324, P49021-4, P07663-1, Q01371, P78714, V5IQ40, Q1K502, P19970-1.

## Funding Statement

J.F.P is funded by the Gerty and Carl Cori Research Investigator Fellowship from the Department of Biochemistry and Molecular Biophysics at Washington University in St. Louis School of Medicine. ETU is funded by the Howard Hughes Medical Institute via award to Carrie Partch.

## Author Contributions

J.F.P and E.T.U conceptualized, wrote, and revised the manuscript.

### Abbreviations

**ARM**: Armadillo (helical repeats)
**bHLH**: Basic Helix Loop Helix
**BMAL**: Basic-helix-loop-helix Arnt-like protein
**CBD**: CRY-binding domain
**CC**: Coiled-coil
**CK1**: Casein Kinase 1
**CKBD**: Casein kinase binding domain
**CLOCK**: Circadian Locomotor Output Cycles Protein Kaput
**CLK**: Circadian Locomotor Output Cycles Protein Kaput, specific to *Drosophila*
**CRY**: Cryptochrome
**DBT**: DOUBLETIME
**DSHCT**: Dob1 Ski2 HelY
**FFD**: FRQ FRH binding Domain
**FRH**: FREQUENCY-Interacting RNA Helicase
**FCD**: FRQ-CK1-interacting domain;
**FRQ**: FREQUENCY
**GSK3**: Glycogen synthase kinase 3
**IDR**: Intrinsically Disordered Region
**IDP**: Intrinsically Disordered Protein
**KOW**: Kyprides, Ouzounis, Woese;
**LCR**: low-complexity region
**LOV**: Light, Oxygen, Voltage
**NTD**: N-terminal Domain
**PAB**: Parp1-binding domain
**PAS**: PER-Arnt-Sim
**PER**: PERIOD
**PHR**: Photolyase Homology Region
**polyQ**: poly-glutamine
**PTM**: Post-translational Modification
**TAD**: Transactivation domain
**WC-1**: White Collar 1
**WC-2**: White Collar 2
**WCC**: White Collar Complex
**WH**: winged helix
**ZnF**: Zinc Finger